# A Vector Matroid-Theoretic Approach in the Study of Structural Controllability Over F(z)

Yupeng Yuan[1], Zhixiong Li, *Member, IEEE*, Malekian Reza, Senior *Member, IEEE*, Yongzhi Chen [4*], Ying Chen[5]

*Abstract*—In this paper, the structural controllability of the systems over F(z) is studied using a new mathematical method-matroids. Firstly, a vector matroid is defined over F(z). Secondly, the full rank conditions of $[sI - A | B]$ ($s \in \wp$) are derived in terms of the concept related to matroid theory, such as rank, base and union. Then the sufficient condition for the linear system and composite system over F(z) to be structurally controllable is obtained. Finally, this paper gives several examples to demonstrate that the married-theoretic approach is simpler than other existing approaches.

## I. Introduction

KALMAN systematically introduced the state space method into systems and control theory, which is the landmark of modern control theory. In the state space method, two important concepts characterizing systems' properties were proposed, namely controllability and observability [1]. After Kalman first proposed the concepts of controllability and observability, many authors [2] studied the subject of linear systems over the field $R$ of real numbers. In short, if the internal movement of all state variables in a system can be affected and controlled by the input, the system can be completely controlled; if all state variables within the system of any form of movement can be completely reflected by the output, the system can be completely observed. It is efficient to use the theory of controllability and observability in real number field for analyzing the controllability and observability jointly determined by system structures and the physical parameter values. The theory has also been successfully applied in many aspects of system analysis and design [2-3]. However, in projects, due to the limitation of experimental conditions, manufacturing process, observational error, and approximate processing of data by human beings, the parameters of a real system are not accurate and even unknown. In such cases, by using only a single index of controllability, it is impossible to decide whether the fact that the system does not satisfy completely controllable conditions is caused by structural reasons or incorrect parameter selection. Therefore, using the system in the real number field is not a direct approach to analyze the structural properties of physical systems [5-6].

Take the bridge network shown in Fig. 1 as an example:

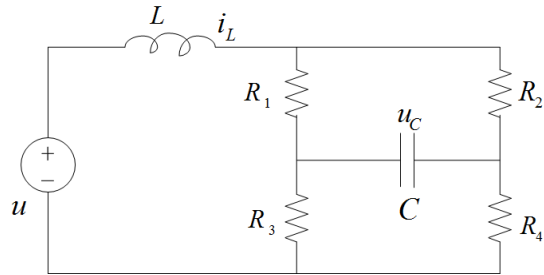

*Figure 1. A bridge network*

Set $x_1 = i_L$, $x_2 = u_C$ as two state variables; then the state equation of the system can be written as:

$$\dot{x}_1 = -\frac{1}{L}\left(\frac{R_1 R_2}{R_1 + R_2} + \frac{R_3 R_4}{R_3 + R_4}\right)x_1 + \frac{1}{L}\left(\frac{R_1}{R_1 + R_2} - \frac{R_3}{R_3 + R_4}\right)x_2 + \frac{1}{L}u$$

$$\dot{x}_2 = \frac{1}{C}\left(\frac{R_2}{R_1 + R_2} - \frac{R_4}{R_3 + R_4}\right)x_1 - \frac{1}{C}\left(\frac{1}{R_1 + R_2} - \frac{1}{R_3 + R_4}\right)x_2 \quad (1)$$

The controllability matrix of the system is

The authors gratefully acknowledge the financial support provided by the National Natural Science Fund grant (51307047 and 51505475) and the self-determined and innovative research funds of Wuhan University of Technology grant 2010-YB-12, Yingcai Project of CUMT, and National Research Foundation, South Africa (RDYR160404161474).

Yupeng Yuan is with the Reliability Engineering Institute, School of Energy and Power Engineering; Key Laboratory of Marine Power Engineering & Technology, (Ministry of Transport); National Engineering Research Center for Water Transport Safety, Wuhan University of Technology, Wuhan 430063,China (Email: ypyuan@whut.edu.cn)

Zhixiong Li is with the School of Mechatronic Engineering, China University of Mining and Technology, Xuzhou 221116, China; and School of Mechanical and Manufacturing Engineering, University of New South Wales, Sydney 2052, Australia (Email: zhixiong.li@ieee.org)

Reza Malekian is with the Department of Electrical, Electronic & Computer Engineering, University of Pretoria, Pretoria 0002, South Africa (Email: reza.malekian@ieee.org)

Yongzhi Chen is with the Reliability Engineering Institute, School of Energy and Power Engineering; Key Laboratory of Marine Power Engineering & Technology, (Ministry of Transport); National Engineering Research Center for Water Transport Safety, Wuhan University of Technology, Wuhan 430063,China (Email: yzchen@whut.edu.cn) [Corresponding author]

Ying Chen is with the Communication staff room of communication department, The Second Artillery Command College, Wuhan, 430063, China (Email: ychen@163.com)

$$[b, Ab] = \begin{bmatrix} \dfrac{1}{L} & -\dfrac{1}{L^2}\left(\dfrac{R_1 R_2}{R_1 + R_2} + \dfrac{R_3 R_4}{R_3 + R_4}\right) \\ 0 & -\dfrac{1}{LC}\left(\dfrac{R_4}{R_3 + R_4} - \dfrac{R_2}{R_1 + R_2}\right) \end{bmatrix} \quad (2)$$

For this system with given structure, only if all physical parameters $L, R_1, R_2, R_3, R_4$ and $C$ are real-valued, $A$ and $b$ are real matrices and the system is a real number system. When $\dfrac{R_4}{R_3 + R_4} \neq \dfrac{R_2}{R_1 + R_2}$, no matter what values $L, R_1, R_2, R_3, R_4$ and $C$ take, it is always true that $\mathrm{rank}[b, Ab] = 2 = n$. Thus, system (1) is controllable everywhere in the real number field. However, when the bridge keeps a dynamic balance, in other words, when $R_1 R_4 = R_2 R_3$, there holds $\dfrac{R_1}{R_1 + R_2} = \dfrac{R_3}{R_3 + R_4}$ and $\dfrac{R_2}{R_1 + R_2} = \dfrac{R_4}{R_3 + R_4}$, then the controllability matrix $[b, Ab]$ is equal to $\begin{bmatrix} \dfrac{1}{L} & -\dfrac{1}{L^2}\left(\dfrac{R_1 R_2}{R_1 + R_2} + \dfrac{R_3 R_4}{R_3 + R_4}\right) \\ 0 & 0 \end{bmatrix}$. Since $\mathrm{rank}[b, Ab] = 1 < n$, system (1) is not controllable in the real number field.

The above example demonstrates that the analytical results of real number system (such as the controllability of $(A, B)$, the observability of $(A^T, C^T)$, and the reducibility of eigenpolynomial $\det(\lambda - A)$ depend on two factors, namely the (physical) structure and values of the physical parameters of a system. However, we cannot determine what the separate role of system structures is.

In order to analyze the system structure, Lin [7] first introduced the concepts of structural controllability (SC) and structured matrix (SM) in 1974 and provided some structural controllability criteria by using graph theory. Later many researchers studied properties of structural controllability. Mayeda studied the controllability of multiple-input multiple-output system [8]. In 1976, the general rank of as tructural matrix is defined by Shield and Pearson[9-10] and the algebraic criterion was obtained. In recent decades, many scholars have studied the structural properties of a system by using the method of algebra or graph theory, and some valuable conclusions were drawn [11-20]. In fact, there exist an equivalence relation between the algebraic criterion and the graph criterion. The algebraic criterion and the graph theory criterion of structural controllability have an equivalent corresponding relationship as described in [21-22] with a rigorous proof. The above studies were based on SM proposed by Lin. Moreover, other forms of parameter matrices have also been proposed and used to study the structural properties of systems. For example, paper [23-26] introduced one-degree polynomial matrix, column-structured matrix (CSM) and mixed matrix (MM). K. Murota first studied the structural controllability of systems by the matroid theory [27-29].He used the union of matroids to analyze the structural controllable conditions of descriptor system whose coefficient matrix is MM. The introduction of matroid further enriched the content of control theory. Although many significant conclusions with regard to structural controllability have been obtained, SM CSM and MM only described certain few types of systems, and the inverse matrix of the full rank square SM, CSM, one-degree polynomial matrix or mixed matrix is generally not SM, CSM, one-degree polynomial matrix or mixed matrix. To overcome this problem, papers [30-36] introduced rational function matrix (RFM)in multi-parameters to describe the coefficient matrices of systems and networks and described the systems and networks on RFM to study their structural properties.

The definition of RFM please see literature [30-36]. For example, if we consider a linear system:

$$\dot{X} = \bar{A}X + \bar{B}U, \quad Y = \bar{C}X + \bar{D}U \quad (3)$$

where $\bar{A}$, $\bar{B}$, $\bar{C}$ and $\bar{D}$ are respectively $\bar{n} \times \bar{n}$, $\bar{n} \times \bar{m}$, $\bar{p} \times \bar{n}$, $\bar{p} \times \bar{m}$ matrices over $F(z)$, the system is called a rational function system (RFS) or a system over $F(z)$.

Since the entries of a rational function matrix are different from those of the SM, which are either fixed zero or independent free parameters, the conception of the RFM is more general. If all the physical parameters of a linear system are viewed as mutually independent parameter variables instead of constant real values, then the system is a linear system in $F(z)$, of which the controllability is irrelevant to system parameter values. If a system is controllable and observable in $F(z)$, i.e., structurally controllable and observable, then it is controllable and observable almost everywhere in parameter space $R^q$, which implies that the system is actually always controllable and observable in the real number field.

Currently, most studies on the structural property of systems in $F(z)$ are based on methods of algebra and graph theory, and many important conclusions have been obtained. However, the algebra or graph criterion for the structural controllability over $F(z)$ is more difficult and complicated than that of the systems which can be described as an SM, especially when the system dimension is very large, since a lot of symbolic operation is required for judging the structural controllability of systems by the criterions in time domain or frequency domain no matter which is obtained based on the state space method or polynomial matrix description method. In this paper, the matroid theory is firstly applied to analyze the system structure over $F(z)$. The equivalent condition that $[sI - \bar{A} \mid \bar{B}]$ is full rank was explored with the concepts in matroid theory, such as base, rank, and union of matroids, and the structural controllability criteria of a system or a parallel composite system are obtained over $F(z)$. The matroids approach is generally easier than the traditional algebraic approach.

## II. PRELIMINARIES

This section presents the basic concepts of matroid-theory. See reference [37] for the complete introduction of matroids[38-41].. We begin with the following definitions.

**Definition 1:** A matroid $M$ is a pair $(E, \mathcal{I})$ where $E$ is a finite set and collection $\mathcal{I} \subseteq 2^E$ is subsets of $E$ such that:

1) $\phi \in \mathcal{I}$;

2) If $I \in \mathcal{I}$ and $I' \subseteq I$ then $I' \in \mathcal{I}$;

3) If $I_1, I_2 \in \mathcal{I}$ and $|I_1| < |I_2|$ then exist $e \in I_2 - I_1$ such that $I_1 \bigcup e \in \mathcal{I}$.

Thus, $(E, \mathcal{I})$ is a matroid which is denoted by $M = M(E, \mathcal{I})$. The entry of collection $\mathcal{I}$ is called an independent set of $M$. Axioms 1)-3) are called independent axioms. Generally, $E = E(M)$ and $\mathcal{I} = \mathcal{I}(M)$ are used to emphasize that $E(M)$ is the set of elements $M$ and $\mathcal{I}(M)$ is the collection of independent sets $M$. If $I \in \mathcal{I}(M)$ and $|I| = k$, $I$ is also called a $k$-independent set of $M$.

**Definition 2:** Let $A$ be an $n \times m$ matrix over $F$ and $E = E(A)$ is a set of labels of the column vectors of $A$. Define $\mathcal{I} \subseteq 2^E$ as a collection: $(E, \mathcal{I})$ is a matroid if and only if the column vectors marked by $X$ are linearly independent in the vector space where $X \in \mathcal{I}$.

**Definition 3:** The maximal independent subset of a matroid $M$ is called a base of $M$. Denote $\mathcal{B}(M)$ the family of all bases of $M$, then $\mathcal{B}(M) = Max(\mathcal{I})$.

**Definition 4**: The function $r_M : 2^E \rightarrow Z$, $r_M(X) = \max\{|I| : I \subseteq X, I \in \mathcal{I}(M)\}$ is called a rank function of the matroid $M$. Value $r_M(E)$ is called the rank of the matroid $M$ and denoted as $r(M)$.

**Definition 5:** Let $M_1, M_2, \cdots, M_n$ be $n$ matroids on the set $E$. Their rank functions are respectively $r_1, r_2, \cdots, r_n$. Let $\mathcal{I} = \{I_1 \bigcup I_2 \bigcup \cdots \bigcup I_n :$ for each $1 \leq i \leq n$, $I_i \in \mathcal{I}(M_i)\}$, then there exists a matroid $M'$ on $E$ such that

1) $\mathcal{I}(M') = \mathcal{I}$;

2) $r_{M'}(X) = \min\{\sum_{i=1}^{n} r_i(Y) + |X - Y| : Y \subseteq X\}$ for any subset $X \subseteq E$.

The matroid $M'$ is usually denoted by $\bigvee_{i=1}^{n} M_i$ and called the union of $M_1, M_2, \cdots, M_n$. With the above definitions, we have the following lemmas.

**Lemma 1:** Let $M(E, \mathcal{I})$ be a matroid and $\mathcal{B} = \mathcal{B}(M)$. $M$ The family $\mathcal{B}$ satisfies

1) $\mathcal{B}$ contains at least one element (which implies that $M$ has at least one base);

2) For $B_1, B_2 \in \mathcal{B}$ and for $x \in B_1 - B_2$, there exists $y \in B_2 - B_1$ such that $(B_1 - x) \bigcup y \in \mathcal{B}$.

**Lemma 2:** Let $M_1$ and $M_2$ be two matroids both on set $E$. Then $\mathcal{B}(M_1 \vee M_2) = \max\{B_1 \bigcup B_2 : B_1 \in \mathcal{B}(M_1), B_2 \in \mathcal{B}(M_2)\}$.

**1. Structural controllability of systems over $F(z)$**

Let $A = \begin{pmatrix} A_1 & A_2 & \cdots & A_n \end{pmatrix}^T$ be matrices over $F(z)$; then we have $M[A] = M[A_1] \vee M[A_2] \vee \cdots \vee M[A_n]$ by the definition of union matroids. To proceed, we need the following concepts.

**Definition 6:** Let $A$ be a matrix over $F(z)$ and $E = E(A)$ be a label set of the column vectors of $A$. $\mathcal{I} \in 2^E$ is defined as a collection: $(E, \mathcal{I})$ is a matroid if and only if the column vectors marked by $X$ are linearly independent in the vector space $V(n, F(z))$, where $X \in \mathcal{I}$. $(E, \mathcal{I})$ is also called vector matroid of $A$ and denoted by $M_{F(z)}[A]$.

**Definition 7:** Let $A$ and $B$ be $n \times n$ and $n \times m$ matrices, respectively. $M[sI - A | B] = (E, \mathcal{I})$ is a vector matroid over $F(z)$. If there exist $n$ column vectors of the matrix $[sI - A | B]$ ($\forall s \in \wp$) whose determinant is a unimodular matrix, the label set $\mathcal{B}$ of $n$ column vectors will be a base of the matroid $M$. $\mathcal{B}$ is also called a unimodular base.

**Lemma 3:** Linear system (3) is structural controllability if and only if $\text{rank}[sI - A | B] = n$, $\forall s \in \wp$.

**Our main results are as follows:**
**Theorem 1:** Consider a linear system (3) over $F(z)$.

Let $\mathcal{A} = [sI - A | B] = \begin{bmatrix} \mathcal{A}_1 \\ \mathcal{A}_2 \end{bmatrix}$. $M[\mathcal{A}_1]$ and $M[\mathcal{A}_2]$ be vector matrices respectively, where $\text{rank}(M[\mathcal{A}_1]) = n_1$, $\text{rank}(M[\mathcal{A}_2]) = n_2$. The system $(A, B)$ is structural controllable if there exist two unimodular bases $B_1$ and $B_2$, satisfy $B_1 \cap B_2 = \phi$ and $n_1 + n_2 = n$ ($\forall s \in \wp$), where $B_1 \in \mathcal{B}(M[\mathcal{A}_1]), B_2 \in \mathcal{B}(M[\mathcal{A}_2])$

**Proof**: Since $\mathcal{A} = [sI - A | B] = \begin{bmatrix} \mathcal{A}_1 \\ \mathcal{A}_2 \end{bmatrix}$, from definition 5 we have $M[\mathcal{A}] = M[\mathcal{A}_1] \vee M[\mathcal{A}_2]$. Additionally, since $B_1$ and $B_2$ are two unimodular bases, $|B_1| = n_1$ and $|B_2| = n_2$ can be established by Definition 7. Since $B_1 \cap B_2 = \phi$ and $|\text{Max}\{B_1 \cup B_2\}| = |B_1| + |B_2| = n_1 + n_2 = n$, where $B_1 \in \mathcal{B}(M[\mathcal{A}_1])$ and $B_2 \in \mathcal{B}(M[\mathcal{A}_2])$, we have $|\mathcal{B}(M[\mathcal{A}_1] \vee M[\mathcal{A}_2])| = n$ from lemma 2. Then $\text{rank}[sI - A | B] = \text{rank}(M[\mathcal{A}]) = \text{rank}(M[\mathcal{A}_1] \vee M[\mathcal{A}_2]) |\mathcal{B}(M[\mathcal{A}_1] \vee M[\mathcal{A}_2])| = n$, so the linear system $(A, B)$ is structural controllable.

**Theorem 2**: Consider a linear system (3) over $F(z)$. Let $\mathcal{A} = [sI - A | B] = \begin{bmatrix} \mathcal{A}_1 \\ \mathcal{A}_2 \\ \vdots \\ \mathcal{A}_n \end{bmatrix}$. $M[\mathcal{A}_1]$, $M[\mathcal{A}_2]$, $\cdots$, $M[\mathcal{A}_n]$ be vector matroids of matrices $\mathcal{A}_1, \mathcal{A}_2, \cdots, \mathcal{A}_n$, where $\text{rank}(M[\mathcal{A}_1]) = n_1$, $\text{rank}(M[\mathcal{A}_2]) = n_2$, $\cdots$, $\text{rank}(M[\mathcal{A}_n]) = n_n$. The system $(A, B)$ is structural controllable if there exist $n$ unimodular bases $B_1, B_2, \cdots, B_n$ satisfy that $B_i \cap B_j = \phi$, where $B_i \in \mathcal{B}(M[\mathcal{A}_i])$, $i, j = 1, 2, \cdots, n$, $i \neq j$ and $n_1 + n_2 + \cdots + n_n = n$, $\forall s \in \wp$.

**Proof**: Since $\mathcal{A} = [sI - A | B] = \begin{bmatrix} \mathcal{A}_1 \\ \mathcal{A}_2 \\ \vdots \\ \mathcal{A}_n \end{bmatrix}$, we have $M[\mathcal{A}] = M[\mathcal{A}_1] \vee M[\mathcal{A}_2] \vee \cdots \vee M[\mathcal{A}_n]$ by Definition 5. It is straightforward to verify that $\mathcal{B}(M_1 \vee M_2 \vee \cdots \vee M_n) = \text{Max}\{B_1 \cup B_2 \cup \cdots \cup B_n\}$ by lemma 2, then $|\mathcal{B}(M_1 \vee M_2 \vee \cdots \vee M_n)| = |\text{Max}\{B_1 \cup B_2 \cup \cdots \cup B_n\}|$. Since $B_1, B_2, \cdots, B_n$ are unimodular bases, where $B_i \in \mathcal{B}(M[\mathcal{A}_i])$ and $B_i \cap B_j = \phi$ ($i \neq j$), we have $|\mathcal{B}(M_1 \vee M_2 \vee \cdots \vee M_n)| = |B_1| + |B_2| + \cdots + |B_n| = n_1 + n_2 + \cdots + n_n = n$. Hence, $\text{rank}[sI - A | B] = \text{rank}(M[\mathcal{A}_1] \vee M[\mathcal{A}_2] \vee \cdots \vee M[\mathcal{A}_n]) = |\mathcal{B}(M_1 \vee M_2 \vee \cdots \vee M_n)| = n$. So the system $(A, B)$ is structural controllable.

Consider two linear subsystems over $F(z)$
$$\Sigma_i : \dot{X}_i = \bar{A}_i X_i + \bar{B}_i u_i, \; y_i = \bar{C}_i X_i + \bar{D}_i u_i \qquad (4)$$
where $\dim \bar{A}_i = n_i \times n_i$, $\dim \bar{B}_i = n_i \times m$, $X_i \in R^{n_i}$, $u_i \in R^m$, $y_i \in R$, $i = 1, 2$. The state space description of the parallel composite system $\Sigma_p$ is:

$$\Sigma_p : \begin{aligned} \begin{bmatrix} \dot{x}_1 \\ \dot{x}_2 \end{bmatrix} &= \begin{bmatrix} \bar{A}_1 & 0 \\ 0 & \bar{A}_2 \end{bmatrix} \begin{bmatrix} x_1 \\ x_2 \end{bmatrix} + \begin{bmatrix} \bar{B}_1 \\ \bar{B}_2 \end{bmatrix} u \\ y &= \begin{bmatrix} \bar{C}_1 & \bar{C}_2 \end{bmatrix} \begin{bmatrix} x_1 \\ x_2 \end{bmatrix} + \begin{bmatrix} \bar{D}_1 + \bar{D}_2 \end{bmatrix} u \end{aligned} \qquad (5)$$

Let $A_i, B_i, C_i$ and $D_i$ be matrices over $F(z)$ ($i = 1, 2$). We have the following conclusions.

**Theorem 3**: A parallel composite system $\Sigma_p$ is structural controllable if $\Sigma_i$ is structural controllable and there exist two unimodular bases $B_1$ and $B_2$ of vector matroids $M[sI - \bar{A}_1 | 0_{n_1 \times n_2} | \bar{B}_1]$ and $M[0_{n_2 \times n_1} | sI - \bar{A}_2 | \bar{B}_2]$, such that $B_1 \cap B_2 = \phi$ ($\forall s \in \wp$).

**Proof**: Since $\Sigma_i$ is structural controllable, we have $\text{rank}[sI - \bar{A}_1 | \bar{B}_1] = \text{rank}(M[sI - \bar{A}_1 | \bar{B}_1]) = n_1$ and $\text{rank}[sI - \bar{A}_2 | \bar{B}_2] = \text{rank}(M[sI - \bar{A}_2 | \bar{B}_2]) = n_2$. Let $\mathcal{A}_1 = [sI - \bar{A}_1 | 0_{n_1 \times n_2} | \bar{B}_1]$ and $\mathcal{A}_2 = [0_{n_2 \times n_1} | sI - \bar{A}_2 | \bar{B}_2]$; then $\mathcal{A} = \begin{bmatrix} \mathcal{A}_1 \\ \mathcal{A}_2 \end{bmatrix} = \begin{bmatrix} sI - \bar{A}_1 | 0_{n_1 \times n_2} | \bar{B}_1 \\ 0_{n_2 \times n_1} | sI - \bar{A}_2 | \bar{B}_2 \end{bmatrix} = [sI - \bar{A} | \bar{B}]$, where $\bar{A} = \begin{bmatrix} \bar{A}_1 & 0 \\ 0 & \bar{A}_2 \end{bmatrix}$ and $\bar{B} = \begin{bmatrix} \bar{B}_1 \\ \bar{B}_2 \end{bmatrix}$. It is true that $M[\mathcal{A}] = M[\mathcal{A}_1] \vee M[\mathcal{A}_2]$ by Definition 5. So $\text{rank}(M[\mathcal{A}]) = \text{rank}(M[\mathcal{A}_1] \vee M[\mathcal{A}_2]) = |\mathcal{B}(M[\mathcal{A}_1] \vee M[\mathcal{A}_2])| = |Max\{B_1 \cup B_2\}|$, where

$B_1 \in \mathcal{B}(M[\mathcal{A}_1])$ and $B_2 \in \mathcal{B}(M[\mathcal{A}_2])$. It is easy to show that $\text{rank}(M[sI - \bar{A}_1 | \bar{B}_1]) = \text{rank}(M[sI - \bar{A}_1 | 0_{n_1 \times n_2} | \bar{B}_1])$ and $\text{rank}(M[sI - \bar{A}_2 | \bar{B}_2]) = \text{rank}(M[0_{n_2 \times n_1} | sI - \bar{A}_2 | \bar{B}_2])$ by linear algebra theory. So $|B_1| = \text{rank}(M[sI - \bar{A}_1 | \bar{B}_1]) = n_1$ and $|B_2| = \text{rank}(M[sI - \bar{A}_2 | \bar{B}_2]) = n_2$. In addition, since $B_1 \cap B_2 = \phi$, we have $\text{rank}(M[\mathcal{A}]) = |Max\{B_1 \cup B_2\}| = |B_1| + |B_2| = n_1 + n_2 = \text{rank}[sI - \bar{A} | \bar{B}]$. Therefore, the parallel composite system $\Sigma_p$ is structural controllable.

We now study the structural controllability of parallel composite system consisting of $N$ subsystems. Let $\Sigma_p$ be such a system whose state space description is:

$$\Sigma_p : \begin{bmatrix} \dot{x}_1 \\ \vdots \\ \dot{x}_N \end{bmatrix} = \begin{bmatrix} \bar{A}_1 & & \\ & \ddots & \\ & & \bar{A}_N \end{bmatrix} \begin{bmatrix} x_1 \\ \vdots \\ x_2 \end{bmatrix} + \begin{bmatrix} \bar{B}_1 \\ \vdots \\ \bar{B}_N \end{bmatrix} u \quad (6)$$

$$y = \begin{bmatrix} \bar{C}_1 & \cdots & \bar{C}_N \end{bmatrix} \begin{bmatrix} x_1 \\ \vdots \\ x_N \end{bmatrix} + \begin{bmatrix} \bar{D}_1 + \cdots + \bar{D}_N \end{bmatrix} u \quad (7)$$

where $\dim \bar{A}_i = n_i \times n_i$, $\dim \bar{B}_i = n_i \times m$, $i = 1, 2, \cdots, N$. Let $A_i$, $B_i$, $C_i$ and $D_i$ are matrices over $F(z)$ $(i = 1, 2, \cdots, n)$. We have:

**Theorem 4**: A parallel composite system $\Sigma_p$ consisting of $N$ subsystems is structural controllable if $\Sigma_i$ are structural controllable and there exist $N$ unimodular bases $B_1, B_2, \cdots, B_N$ of vector matroids $M[sI - \bar{A}_1 | 0_{n_1 \times (n_2 + n_3 + \cdots + n_N)} | \bar{B}_1]$ $M[0_{n_2 \times n_1} | sI - \bar{A}_2 | 0_{n_2 \times (n_3 + n_4 + \cdots + n_N)} | \bar{B}_2]$, $\cdots$, $M[0_{n_N \times (n_1 + n_2 + \cdots + n_{N-1})} | sI - \bar{A}_N | \bar{B}_N]$ satisfying that $B_i \cap B_j = \phi$, $i, j = 1, 2, \cdots, N$, $i \neq j$, $\forall s \in \wp$.

**Proof**: Since $\Sigma_i$ are structural controllable, we have $\text{rank}[sI - \bar{A}_1 | \bar{B}_1] = \text{rank}(M[sI - \bar{A}_1 | \bar{B}_1]) = n_1$, $\text{rank}[sI - \bar{A}_2 | \bar{B}_2] = \text{rank}(M[sI - \bar{A}_2 | \bar{B}_2]) = n_2$, $\cdots$, $\text{rank}[sI - \bar{A}_N | \bar{B}_N] = \text{rank}(M[sI - \bar{A}_N | \bar{B}_N]) = n_N$. Let $\mathcal{A}_1 = [sI - \bar{A}_1 | 0_{n_1 \times (n_2 + n_3 + \cdots + n_N)} | \bar{B}_1]$, $\mathcal{A}_2 = [0_{n_2 \times n_1} | sI - \bar{A}_2 | 0_{n_2 \times (n_3 + n_4 + \cdots + n_N)} | \bar{B}_2]$, $\cdots$, $\mathcal{A}_N = [0_{n_N \times (n_1 + n_2 + \cdots + n_{N-1})} | sI - \bar{A}_N | \bar{B}_N]$; then

$$\mathcal{A} = \begin{bmatrix} \mathcal{A}_1 \\ \mathcal{A}_2 \\ \vdots \\ \mathcal{A}_N \end{bmatrix} = \begin{bmatrix} sI - \bar{A}_1 | 0_{n_1 \times (n_2 + n_3 + \cdots + n_N)} | \bar{B}_1 \\ 0_{n_2 \times n_1} | sI - \bar{A}_2 | 0_{n_2 \times (n_3 + n_4 + \cdots + n_N)} | \bar{B}_2 \\ \cdots \\ 0_{n_N \times (n_1 + n_2 + \cdots + n_{N-1})} | sI - \bar{A}_N | \bar{B}_N \end{bmatrix} \quad (8)$$

$$= [sI - \bar{A} | \bar{B}]$$

where

$$\bar{A} = \begin{bmatrix} \bar{A}_1 & & 0 \\ & \ddots & \\ 0 & & \bar{A}_N \end{bmatrix}, \quad \bar{B} = \begin{bmatrix} \bar{B}_1 \\ \vdots \\ \bar{B}_N \end{bmatrix} \quad (9)$$

We have $M[\mathcal{A}] = M[\mathcal{A}_1] \vee M[\mathcal{A}_2] \vee \cdots \vee M[\mathcal{A}_N]$ by Definition 5.

So
$\text{rank}(M[\mathcal{A}]) = \text{rank}(M[\mathcal{A}_1] \vee M[\mathcal{A}_2] \vee \cdots \vee M[\mathcal{A}_N])$
$= |Max\{B_1 \cup B_2 \cup \cdots \cup B_N\}|$
$= |\mathcal{B}(M[\mathcal{A}_1] \vee M[\mathcal{A}_2] \vee \cdots \vee M[\mathcal{A}_N])|$

where $B_1 \in \mathcal{B}(M[\mathcal{A}_1])$, $B_2 \in \mathcal{B}(M[\mathcal{A}_2])$, $\cdots$, $B_N \in \mathcal{B}(M[\mathcal{A}_N])$

Since
$\text{rank}(M[sI - \bar{A}_1 | \bar{B}_1]) = \text{rank}(M[sI - \bar{A}_1 | 0_{n_1 \times (n_2 + n_3 + \cdots + n_N)} | \bar{B}_1])$
$\text{rank}(M[sI - \bar{A}_2 | \bar{B}_2]) = \text{rank}(M[0_{n_2 \times n_1} | sI - \bar{A}_2 | 0_{n_2 \times (n_3 + n_4 + \cdots + n_N)} | \bar{B}_2])$
$\cdots$
$\text{rank}(M[sI - \bar{A}_N | \bar{B}_N]) = \text{rank}(M[0_{n_N \times (n_1 + n_2 + \cdots + n_{N-1})} | sI - \bar{A}_N | \bar{B}_N])$

So
$|B_1| = \text{rank}(M[sI - \bar{A}_1 | \bar{B}_1]) = n_1$, $|B_2| = \text{rank}(M[sI - \bar{A}_2 | \bar{B}_2]) = n_2$, $\cdots$, $|B_N| = \text{rank}(M[sI - \bar{A}_N | \bar{B}_N]) = n_N$
and, in addition,
$B_i \cap B_j = \phi$, $i, j = 1, 2, \cdots, N$, $i \neq j$
$\text{rank}(M[\mathcal{A}]) = |Max\{B_1 \cup B_2 \cup \cdots \cup B_N\}|$
$= |B_1| + |B_2| + \cdots + |B_N|$
$= n_1 + n_2 + \cdots + n_N$
$= \text{rank}[sI - \bar{A} | \bar{B}]$

Therefore, the parallel composite system $\Sigma_p$ is structural controllable.

III. EXAMPLE

**Example 1**: Consider a parallel composite system [42-45]

shown in figure 2.

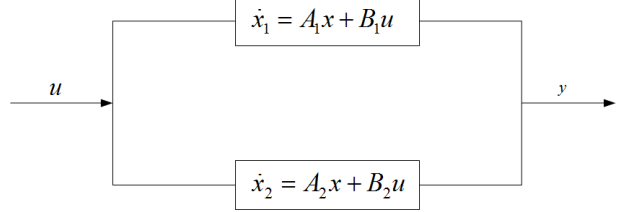

Figure 2 A parallel composite system

where

$$A_1 = \begin{bmatrix} z_1 & 1 \\ 0 & z_2 \end{bmatrix}, \quad B_1 = \begin{bmatrix} 0 & 0 \\ z_3 & 1 \end{bmatrix}, \quad A_2 = \begin{bmatrix} 1 & 1 & 0 \\ 0 & 0 & 1 \\ 1 & 0 & 0 \end{bmatrix},$$

$$B_2 = \begin{bmatrix} z_1 & 0 \\ 0 & 1 \\ 0 & 0 \end{bmatrix}$$ are matrices over $F(z)$.

The controllability matrix of subsystem $\Sigma_1 : \dot{x}_1 = A_1 x_1 + B_1 u$ of parallel composite system is

$$[B_1 \quad A_1 B_1] = \begin{bmatrix} 0 & 0 & z_3 & 1 \\ z_3 & 1 & z_2 z_3 & z_2 \end{bmatrix} \quad (10)$$

Since $\text{rank}[B_1 \quad A_1 B_1] = 2$, the subsystem $\Sigma_1 : \dot{x}_1 = A_1 x_1 + B_1 u$ is structural controllable. Similarly, the controllability matrix of subsystem $\Sigma_2 : \dot{x}_2 = A_2 x_2 + B_2 u$ is

$$[B_2 \quad A_2 B_2 \quad A_2^2 B_2] = \begin{bmatrix} z_1 & 0 & z_1 & 1 & z_1 & 1 \\ 0 & 1 & 0 & 0 & z_1 & 0 \\ 0 & 0 & z_1 & 0 & z_1 & 0 \end{bmatrix} \quad (11)$$

Then, $\text{rank}[B_2 \quad A_2 B_2 \quad A_2^2 B_2] = 3$, so the subsystem

$$\Sigma_p (p=1,2): \begin{bmatrix} \dot{x}_1 \\ \dot{x}_2 \end{bmatrix} = \begin{bmatrix} A_1 & 0 \\ 0 & A_2 \end{bmatrix} \begin{bmatrix} x_1 \\ x_2 \end{bmatrix} + \begin{bmatrix} B_1 \\ B_2 \end{bmatrix} u \quad (12)$$

Let $A = \begin{bmatrix} A_1 & 0 \\ 0 & A_2 \end{bmatrix}$ and $B = \begin{bmatrix} B_1 \\ B_2 \end{bmatrix}$; then

$$[sI - A | B] = \begin{bmatrix} sI - A_1 | 0 | B_1 \\ 0 | sI - A_2 | B_2 \end{bmatrix} =$$

$$\begin{array}{ccccccc} a_1 & a_2 & a_3 & a_4 & a_5 & a_6 & a_7 \\ \begin{bmatrix} s-z_1 & -1 & 0 & 0 & 0 & 0 & 0 \\ 0 & s-z_2 & 0 & 0 & 0 & z_3 & 1 \\ 0 & 0 & s-1 & -1 & 0 & z_1 & 0 \\ 0 & 0 & 0 & s & -1 & 0 & 1 \\ 0 & 0 & -1 & 0 & s & 0 & 0 \end{bmatrix} \end{array} \quad (13)$$

where $E = \{a_1, a_2, a_3, a_4, a_5, a_6, a_7\}$ is the label set of the column vector of the matrix $[sI - A | B]$. Let

$$\mathcal{A}_1 = [sI - A_1 | 0 | B_1], \quad \mathcal{A}_2 = [0 | sI - A_2 | B_2] \quad (14)$$

then $M[\mathcal{A}_1]$ and $M[\mathcal{A}_2]$ are two vector matroids over $F(z)$. So

$$\text{rank}(M[\mathcal{A}_1]) = n_1 = 2, \quad \text{rank}(M[\mathcal{A}_2]) = n_2 = 3 \quad (15)$$

$B_1 = \{a_2, a_6\}$ and $B_2 = \{a_3, a_5, a_7\}$ are unimodular bases of vector matroid $M[\mathcal{A}_1]$ and $M[\mathcal{A}_2]$ respectively by linear algebra theory and the definition of matroid base. Moreover, since

$$B_1 \cap B_2 = \{a_2, a_6\} \cap \{a_3, a_5, a_7\} = \phi \quad \text{and}$$

$n_1 + n_2 = 2 + 3 = 5$,

The parallel composite system is structural controllable by Theorem 3.

**Example 2:** The classical double inverted pendulum system shown in figure 3 is composed of a pair of rigid, homogenous pendulum rods which are interconnected in a joint, and one of them is attached to a stable mechanism (cart) which allows for movement alongside a single axis. To simplify the system, we ignore air resistance and all sorts of friction.

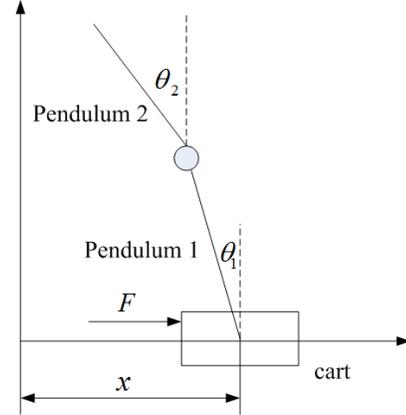

Figure 3 Schematic diagram of double inverted pendulum system

Here $\theta_1$ and $\theta_2$ respectively represent the angles of first and second pendulums with respect to the vertical axis, $x$ denotes the position of the cart, $M$ is the mass of the cart, $m_1$ is the mass of the first link, $m_2$ represents the mass of second link, $m_3$ is the mass of the hinge and encoder, $l_1$ and $l_2$ denote the length of the first and second pendulums, respectively. State variables are taken as follows:

$x_1 = x$, $x_2 = \theta_1$, $x_3 = \theta_2$, $x_4 = \dot{x}$, $x_5 = \dot{\theta}_1$, $x_6 = \dot{\theta}_2$

Let $Z = (z_1, z_2, z_3, z_4, z_5) = (m_1, m_2, m_3, l_1, l_2)$ be five independent physical parameters. The state space equation of double inverted pendulum system can then be written as:

$$\dot{x} = Ax + Bu \quad (16)$$

where

$$A = \begin{bmatrix} 0 & 0 & 0 & 1 & 0 & 0 \\ 0 & 0 & 0 & 0 & 1 & 0 \\ 0 & 0 & 0 & 0 & 0 & 1 \\ 0 & 0 & 0 & 0 & 0 & 0 \\ 0 & K_{12} & K_{13} & 0 & 0 & 0 \\ 0 & K_{22} & K_{23} & 0 & 0 & 0 \end{bmatrix}, \quad B = \begin{bmatrix} 0 \\ 0 \\ 0 \\ 1 \\ K_{17} \\ K_{27} \end{bmatrix} \quad (17)$$

$$\frac{3g(z_1 + 2z_2 + 2z_3)}{z_4(4z_1 + 3z_2 + 12z_3)} = K_{12} \quad (18)$$

$$-\frac{9z_2 g}{2z_4(4z_1 + 3z_2 + 12z_3)} = K_{13} \quad (19)$$

$$-\frac{9g(z_1 + 2z_2 + 2z_3)}{2z_5(4z_1 + 3z_2 + 12z_3)} = K_{22} \quad (20)$$

$$-\frac{3g(z_1 + 3z_2 + 3z_3)}{z_5(4z_1 + 3z_2 + 12z_3)} = K_{23} \quad (21)$$

$$\frac{3(2z_1 + z_2 + 4z_3)}{2z_4(4z_1 + 3z_2 + 12z_3)} = K_{17} \quad (22)$$

$$-\frac{3z_1}{2z_5(4z_1 + 3z_2 + 12z_3)} = K_{27} \quad (23)$$

Let

$$\mathcal{A}_1 = \begin{bmatrix} s & 0 & 0 & -1 & 0 & 0 & 0 \\ 0 & s & 0 & 0 & -1 & 0 & 0 \end{bmatrix} \quad (24)$$

$$\mathcal{A}_2 = \begin{bmatrix} 0 & 0 & s & 0 & 0 & -1 & 0 \\ 0 & 0 & 0 & s & 0 & 0 & 1 \end{bmatrix} \quad (25)$$

$$\mathcal{A}_3 = \begin{bmatrix} 0 & -K_{12} & -K_{13} & 0 & s & 0 & K_{17} \\ 0 & -K_{22} & -K_{23} & 0 & 0 & s & K_{27} \end{bmatrix} \quad (26)$$

We have

$$\mathcal{A} = [sI - A \mid B] = \begin{bmatrix} \mathcal{A}_1 \\ \mathcal{A}_2 \\ \mathcal{A}_3 \end{bmatrix} =$$

$$\begin{array}{ccccccc} a_1 & a_2 & a_3 & a_4 & a_5 & a_6 & a_7 \end{array} \quad (26)$$

$$\begin{bmatrix} s & 0 & 0 & -1 & 0 & 0 & 0 \\ 0 & s & 0 & 0 & -1 & 0 & 0 \\ 0 & 0 & s & 0 & 0 & -1 & 0 \\ 0 & 0 & 0 & s & 0 & 0 & 1 \\ 0 & -K_{12} & -K_{13} & 0 & s & 0 & K_{17} \\ 0 & -K_{22} & -K_{23} & 0 & 0 & s & K_{27} \end{bmatrix}$$

where $E = \{a_1, a_2, a_3, a_4, a_5, a_6, a_7\}$ is the label set of the column vectors of matrix $\mathcal{A}$. Then $M[\mathcal{A}_1]$, $M[\mathcal{A}_2]$ and $M[\mathcal{A}_3]$ are all vector matroids over $F(z)$, and $M[\mathcal{A}] = M[\mathcal{A}_1] \vee M[\mathcal{A}_2] \vee M[\mathcal{A}_3]$. $B_1 = \{a_4, a_5\}$, $B_2 = \{a_6, a_7\}$ and $B_3 = \{a_2, a_3\}$ are unimodular bases of vector matroid $M[\mathcal{A}_1]$, $M[\mathcal{A}_2]$ and $M[\mathcal{A}_3]$ respectively by the definition of matroid base. Moreover, since

$$\text{rank} M[\mathcal{A}_1] = |B_1| = 2, \quad \text{rank} M[\mathcal{A}_2] = |B_2| = 2,$$

$$\text{rank} M[\mathcal{A}_3] = |B_3| = 3$$

and $|B_1| + |B_2| + |B_3| = 6 = n$, $B_i \cap B_j = \phi$, $i, j = 1, 2, 3$, $i \neq j$, $\forall s \in \wp$, the double inverted pendulum system is structural controllable by Theorem 4.

IV. CONCLUSION

The structural controllability of linear systems was studied with different mathematical methods and matrix structures, and some structural controllability criteria were derived. In this paper, the structural property has been studied using the concept of matroids over $F(z)$. Then, the structural controllability criteria have been obtained for linear systems and parallel composite systems. A novel approach to study the structural properties has also been provided. We have shown that the approach presented in this paper is substantially simpler than any other existing ones.